\documentclass[11pt]{article}

\usepackage{latexsym,cmmib57}
\usepackage{amsthm}
\usepackage{amsmath}
\usepackage[psamsfonts]{amssymb,amsfonts}
\usepackage{graphicx}
\newtheorem{definition}{Definition}
\newtheorem{theorem}{Theorem}
\newtheorem{lemma}{Lemma}

\newcounter{algocnt}
\newenvironment{algolist}[1]{%
    \begin{list}{\thealgocnt}
    {\parsep 0pt\usecounter{algocnt}\setcounter{algocnt}{0}\renewcommand{\thealgocnt}{{#1}\arabic{algocnt}}
    \setlength{\rightmargin}{0in}
    \settowidth{\leftmargin}{{#1}999}\addtolength{\leftmargin}{\labelsep}}}{\end{list}}
\newcommand{\algotab}[0]{\hspace*{2\labelsep}}
\newcommand{\mt}[0]{\algotab\ }
\newcommand{\mtt}[0]{\algotab\ \algotab\ }
\newcommand{\mttt}[0]{\algotab\ \algotab\ \algotab\ }

\newcommand{\Algo}[1]{\textsc{#1}}
\newcommand{\setto}{\leftarrow}

\newsavebox{\fmbox}

\newcommand{\size}{n}
\newcommand{\wt}{a}
\newcommand{\wthreshold}{\alpha}
\newcommand{\pre}{f}
\newcommand{\lladder}{l}
\newcommand{\rladder}{r}
\newcommand{\rpair}[1]{\mathsf{right}(#1)}
\newcommand{\lpair}[1]{\mathsf{left}(#1)}

\begin{document}
\title{A linear-time algorithm for 
	finding 
	the longest segment which scores above a given threshold}
\author{Mikl\'os Cs\H{u}r\"os\thanks{
	Universit\'e de Montr\'eal,	
	Department of Computer Science and Operations Research,
	CP 6128 succ.\ Centre-Ville, Montr\'eal, Qu\'e., H3C~3J7, Canada.
	WWW: http://www.iro.umontreal.ca/\textasciitilde{}csuros/}.}

\maketitle
	
\paragraph{Abstract.}
This paper describes a linear-time algorithm that finds the
longest stretch in a sequence of real numbers (``scores'')
in which the sum exceeds an input parameter. The algorithm
also solves the problem of finding the longest interval in
which the average of the scores is above a fixed threshold.
The problem originates from molecular sequence analysis: for
instance, the algorithm can be employed to identify long
GC-rich regions in DNA sequences. The algorithm can also be
used to trim low-quality ends of shotgun sequences in a
preprocessing step of whole-genome assembly.

\section{Introduction} 
Let
$\wt_1,\dotsc,\wt_{\size}$
be an arbitrary sequence of 
real numbers with~$\size>0$.
The {\em segment} $[i,j]$ 
for $1\le i\le j\le n$ 
is the interval $\{i,i+1,\dotsc,j\}$;
its {\em score} is 
$
\wt(i,j)=\wt_i+\wt_{i+1}+\dotsm+\wt_j$.
This paper's central problem 
is the following.
Given a score threshold~$\wthreshold$,
find a segment~$[i,j]$ that has maximum length
$(j-i+1)$ among those with $\wt(i,j)\ge\wthreshold$.

Similar segmentation questions are
encountered 
in statistical change-point estimation~\cite{FuCurnow}, 
with applications in 
various areas including 
molecular biology~\cite{AugerLawrence,segment-sets,Karlin.scores}.
A number of related problems 
can be solved with efficient algorithms. 
Jon Bentley's classic ``programming pearl''
finds a segment with maximum score 
in~$O(\size)$ time~\cite{subsum}.
Cs\H{u}r\"os~\cite{segment-sets} 
solves the more general problem of finding 
a $k$-set of segments with maximum total score 
in $O(\size\min\{k,\log \size\})$ time 
and~$O(\size)$ space.
Huang~\cite{subsum.Huang} 
reports a simple linear-time algorithm for 
the dual of our problem, namely,
that of finding 
a a segment that has maximum score among 
those longer than a given threshold.
An algorithm  of Lin {\em et al.}~\cite{segment.avg} 
finds such a segment in~$O(\size)$ time, 
when in addition to 
a lower bound on the segment length, an upper bound is also 
imposed.  

In some situations, it may be interesting 
to evaluate a segment~$[i,j]$ by its {\em average score}
$\wt(i,j)/(j-i+1)$. 
Lin {\em et al.}~\cite{segment.avg} devised an algorithm
that finds the segment with maximum average score
among those longer than~$L$, in $O(\size\log L)$ time. 
Goldwasser {\em et al.}~\cite{Goldwasser.avg} 
give a faster algorithm for the same problem that 
runs in~$O(\size)$ time irrespective of~$L$. 
This paper's techniques lead to
an~$O(\size)$-time algorithm for the dual problem;
namely, that of finding the longest segment with average score
above a bound~$\wthreshold$. 
This latter result is particularly relevant in molecular 
sequence segmentation. For instance,
our algorithm can be employed 
to identify the longest contiguous region in a DNA sequence
with a GC-content (relative frequency of guanine and cytosine) 
above a cutoff level. 
The search for GC-rich and GC-poor 
regions in DNA 
is one of the main practical motivations 
behind the algorithms of~\cite{segment-sets,Goldwasser.avg,subsum.Huang,segment.avg}.

Whole-genome shotgun assembly programs also often 
need to compute long segments with high average scores.
In shotgun sequencing, 
the sequence of a long DNA molecule is 
computed from the sequences of randomly sampled 
short fragments~\cite{Sequencing.review}, called 
the {\em shotgun sequences}.
The shotgun sequences are typically 
delivered together with position-specific error 
probabilities~\cite{Phred.error}
to the assembly software. 
In a preprocessing phase, many assembly 
programs trim the shotgun sequences by removing the extremities 
with high sequencing error. 
It is important to trim the sequences only as much as
is absolutely necessary. Small error levels can be tolerated and 
even corrected, while the assembly's quality is 
ultimately determined by its length.
The shotgun sequence 
trimming problem is defined as follows. 
Given a DNA sequence $s_1s_2\dotsm s_n$ and 
position-specific error probabilities 
$e_1,e_2,\dotsc, e_n$, 
find the longest contiguous substring $s_i\dotsc s_j$ 
such that its average error $(e_i+e_{i+1}+\dotsc+e_j)/(j-i+1)$
falls below a user-specified threshold~$E$. Clearly, by setting 
$a_k=1-e_k$, we 
can look for the longest segment 
for which the average score is above~$(1-E)$ by using 
the techniques in this paper. 
Existing assembly programs
trim heuristically using variations 
of a sliding window technique, 
without guarantees of length optimality.
(They rely on the fact that the error
probabilities in chain-termination sequencing 
are usually high at the extremities and low in the middle, 
and essentially assume a unimodal function.)
The assembly program Arachne~\cite{Arachne}, for instance, 
purposely looks for the longest segment with 
an average error below a threshold, but 
closer inspection of the source code 
reveals that the implemented algorithm
is not guaranteed to find an optimal segment for 
all error probabilities.
%(David Jaffe, personal communication). 

\section{Algorithm}
%This section describes a linear-time algorithm that
%finds the longest segment that scores above a 
%fixed parameter~$\wthreshold$.
Define the {\em prefix score} $\pre_j=\wt(1,j)$
for all $j=1,\dotsc,\size$,
and let $\pre_0=0$. 
Obviously, $\wt(i,j)=\pre_j-\pre_{i-1}$,
and thus we are looking for the longest segment
$[i,j]$ with $\pre_j\ge\pre_{i-1}+\wthreshold$.
Now, 
let $0\le i^*<j^*\le \size$ be such 
that $\pre_{j^*}\ge\pre_{i^*}+\wthreshold$ and 
$(j^*-i^*)$ is maximal.
Clearly, $[i^*+1,j^*]$ is the longest segment
with $a(i^*+1,j^*)\ge \wthreshold$. 
\begin{lemma}\label{lm:peaks}
Let $0\le i^*<j^*\le \size$ be such 
that $\pre_{j^*}\ge\pre_{i^*}+\wthreshold$ and 
$(j^*-i^*)$ is maximal.
\begin{subequations}
If $i^*>0$, then
\begin{equation}\label{eq:peaks.left}
\pre_{i^*} <\pre_0,\pre_1,\dotsc,\pre_{i^*-1}.
\end{equation}
If $j^*<\size$, then
\begin{equation}\label{eq:peaks.right}
\pre_{j^*}>\pre_{\size},\pre_{\size-1},\dotsc,\pre_{j^*+1}.
\end{equation}
\end{subequations}
\end{lemma}
\begin{proof}
We prove Eq.~\eqref{eq:peaks.left}.
For the sake of contradiction, assume that there exists 
such an $i<i^*$ that 
$\pre_i\le\pre_{i^*}$. 
Then $j^*-i>j^*-i^*$, yet
$\pre_{j^*}\ge\pre_{i^*}+\wthreshold\ge\pre_i+\wthreshold$.
Eq.~\eqref{eq:peaks.right} is proven analogously.
\end{proof}

\begin{definition}\label{def:peaks}
Define the {\em left sequence of minima}
$0=\lladder_1<\lladder_2<\dotsb\lladder_k\le \size$ 
by $\lladder_1=0$ and 
$\lladder_j=\min\{i\colon\lladder_{j-1}<i\le\size, \pre_i<\pre_{\lladder_{j-1}}\}$.
Define the {\em right sequence of maxima}
$\size=\rladder_1>\rladder_2>\dotsb>\rladder_m\ge 0$
by 
$\rladder_1=\size$ and
$\rladder_j=\max\{i\colon 0\le i<\rladder_{j-1}, \pre_i>\pre_{\rladder_{j-1}}\}$.
\end{definition}

\begin{figure}
\centerline{\includegraphics[height=0.16\textheight]{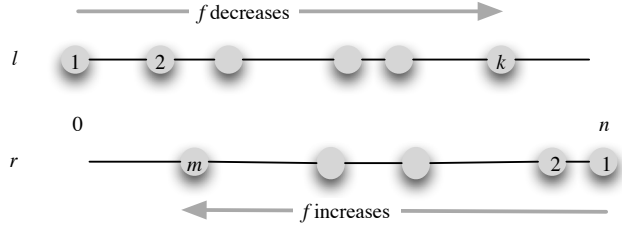}}
\caption{Left sequence of minima $\{l_i\}$
and right sequence of maxima $\{r_j\}$. 
}\label{fig:ladder}
\end{figure}

Figure~\ref{fig:ladder} illustrates 
these notions.
By definition, 
the left sequence of minima is 
sorted in decreasing order of prefix scores:
\begin{subequations}
\begin{gather}\label{eq:peaks.left.decreasing}
\pre_{\lladder_1}>\pre_{\lladder_2}>\dotsb >\pre_{\lladder_k}.\\
\intertext{Similarly,} 
%the right sequence of maxima 
%is sorted in increasing order of prefix scores:}
\label{eq:peaks.right.increasing}
\pre_{\rladder_1}<\pre_{\rladder_{2}}<\dotsb<\pre_{\rladder_m}.
\end{gather}
\end{subequations}

By Lemma~\ref{lm:peaks}, we can restrict our attention to
segments~$[i,j]$ where $i\in\{\lladder_1\dotsc,\lladder_k\}$ and
$j\in\{\rladder_1,\dotsc,\rladder_m\}$. 
Equations~\eqref{eq:peaks.left.decreasing} 
and~\eqref{eq:peaks.right.increasing} imply the following lemmas.

\begin{lemma}\label{lm:peaks.right.all}
Let $i\in\{0,\dotsc,\size\}$.
If $\pre_i+\wthreshold\le\pre_{\rladder_j}$ for some $j$, then 
$\pre_i+\wthreshold\le\pre_{\rladder_j'}$ for all $j'\ge j$.
\end{lemma}
\begin{lemma}\label{lm:peaks.left.all}
Let $j\in\{0,\dotsc,\size\}$.
If $\pre_{\lladder_i}+\alpha\le\pre_j$, then 
$\pre_{\lladder_{i'}}+\alpha\le\pre_j$ for all $i'\ge i$. 
\end{lemma}
In view of these lemmas, we can define the following values.

\begin{definition}\label{def:pairs}
For all $i=1,\dotsc,k$ define 
$\rpair{i}$
by 
$
\rpair{i} = \min \{j\colon f_{\rladder_j}\ge f_{\lladder_i}+\wthreshold\}
$.
Let $\rpair{i}=m+1$ if $f_{\lladder_i}+\wthreshold>f_{\rladder_j}$ for all $j$.

For all $j=1,\dotsc,m$, define 
$\lpair{j}$ 
by 
$
\lpair{j} = \min\{i\colon f_{\lladder_i}+\wthreshold\le f_{\rladder_j}\};
$
let $\lpair{j}=k+1$ if $f_{\lladder_i}+\wthreshold>f_{\rladder_j}$ for all $i$.
\end{definition}
By Lemmas \ref{lm:peaks}, \ref{lm:peaks.right.all} and~\ref{lm:peaks.left.all},
for every $i=1,\dotsc, k$, 
the longest segment $[\lladder_i+1,j]$ which scores above~$\wthreshold$ 
has $j=\rladder_{\rpair{i}}$,
unless $\rpair{i}=m+1$ or $\rladder_{\rpair{i}}\le\lladder_i$, 
in which case there is no suitable segment with left endpoint 
at $\lladder_i+1$.
Similarly, $\lpair{j}$ gives the left endpoint $i=\lladder_{\lpair{j}}$
for the longest segment $[i+1,\rladder_j]$ that scores above the threshold,
unless $\lpair{j}=k+1$ or $\lladder_{\lpair{j}}\ge\rladder_j$, 
in which case there is no suitable segment with right
endpoint~$\rladder_j$. 
Lemmas~\ref{lm:peaks.right.all} and~\ref{lm:peaks.left.all}
imply the following property.
\begin{lemma}\label{lm:pairs.ordered}
For all $1\le i<i'\le k$, $\rpair{i}\ge\rpair{i'}$. 
For all $m\ge j>j'\ge 1$, $\lpair{j}\le\lpair{j'}$.
\end{lemma}
Now, the best segment is the longest valid segment in
the set 
\[
\bigl\{[\lladder_i+1,\rladder_{\rpair{i}}]\colon i=1,\dotsc,k\bigr\}
\cup
\bigl\{[\lladder_{\lpair{j}}+1,\rladder_j]\colon j=1,\dotsc, m\bigr\}.
\]
In fact, it suffices to consider only one of the two sets 
since for the longest segment $[\lladder_{i^*}+1,\rladder_{j^*}]$,
$\lpair{j^*}=i^*$ and $\rpair{i^*}=j^*$. 

The following algorithm solves the original problem.
\begin{center}
%\begin{fmpage}{0.95\textwidth}
\begin{algolist}{}
\item \textbf{Algorithm} \Algo{LongestSegment}
\item \textbf{Input}: scores $a_i\colon i=1,\dotsc,\size$; threshold $\alpha$
\item \textbf{Output}: longest segment that scores above $\alpha$, or $\mathsf{nil}$
	if no segment score exceeds~$\alpha$
\item Set $\pre_0\setto 0$; \textbf{for} $i\setto 1,\dotsc, \size$ 
	\textbf{do} $\pre_i\setto\pre_{i-1}+\wt_i$\label{line:prefix}
\item Set $k\setto 1, \lladder_1\setto 0$\label{line:lmin.1}
\item \textbf{for} $i\setto 1,\dotsc, \size$ 
	\textbf{do}
	\textbf{if} $\pre_i<\pre_{\lladder_k}$
	\textbf{then} $k\setto k+1, \lladder_k\setto i$\label{line:lmin.2}
\item Set $m\setto 1, \rladder_1\setto\size$\label{line:rmax.1}
\item \textbf{for} $j\setto\size,\dotsc, 1$
	\textbf{do}
	\textbf{if} $\pre_j>\pre_{\rladder_m}$
	\textbf{then} $m\setto m+1, \rladder_m\setto j$\label{line:rmax.2}
\item Set $\mathsf{max}\setto 0, \mathsf{segment}\setto\mathsf{nil}$\label{line:seg.init}
\item Set $i\setto 1, j\setto m$\label{line:indexes.init}
\item \textbf{while} $i\le k$ and $j\ge 1$ \textbf{do}\label{line:loop.outer}
\item\mt\ \textbf{while} $i\le k$ and $\pre_{\lladder_i}+\wthreshold>\pre_{\rladder_j}$
	\textbf{do} $i\setto i+1$ \label{line:skip.left}
\item\mt\ \textbf{if} $i\le k$ \textbf{then}\label{line:test.left}
\item\mtt\ \textbf{while} $j\ge 1$ 
	and $\pre_{\lladder_i}+\wthreshold\le\pre_{\rladder_j}$ \textbf{do}
		\label{line:loop.inner}
\item\mttt\ \textbf{if} $\rladder_j-\lladder_i>\mathsf{max}$ \textbf{then}
	$\mathsf{max}\setto \rladder_j-\lladder_i,
	\mathsf{segment}\setto[\lladder_i+1,\rladder_j]$\label{line:seg.test}
\item\mttt\ Set $j\setto j-1$\label{line:skip.right}
\item\mtt\ \textbf{end while}\label{line:loop.inner.end}
\item\mt\ \textbf{end if} 
\item \textbf{end while}\label{line:loop.outer.end}
\item \textbf{return} $\mathsf{segment}$
\end{algolist}
%\end{fmpage}
\end{center}
\begin{theorem}
Algorithm \Algo{LongestSegment} finds the longest segment
which scores above the threshold~$\wthreshold$ in $O(\size)$
time.
\end{theorem}
\begin{proof}
Line~\ref{line:prefix} calculates the prefix sums~$\pre_i$ 
in~$O(\size)$ time. Lines~\ref{line:lmin.1}--\ref{line:lmin.2}
compute the left sequence of minima, and 
Lines~\ref{line:rmax.1}--\ref{line:rmax.2} 
compute the right sequence of maxima, in~$O(\size)$ time.
Lines~\ref{line:loop.outer}--\ref{line:loop.outer.end}
are executed in $O(k+m)=O(n)$
time, since the loops of Lines~\ref{line:skip.left}
and~\ref{line:skip.right} increase~$i$ and decrease~$j$ by one, 
respectively. 

Line~\ref{line:seg.init} initializes the 
structures for tracking the best segment: 
$\mathsf{max}$ stores the length of
the longest segment found and $\mathsf{segment}$
is the best segment. 
In Lines~\ref{line:indexes.init}--\ref{line:loop.outer.end},
the algorithm goes through 
pairs $(i,j)$ where $i=\lpair{j}$. 
More precisely, 
the algorithm's correctness follows from the invariant that 
$i=k+1$ or $i=\lpair{j}$ holds in Line~\ref{line:test.left}. 
Subsequently, as long as the \textbf{while} loop's condition 
in Line~\ref{line:loop.inner}
is true, $i=\lpair{j}$ holds.
As discussed, one of the segments 
$[\lladder_{\lpair{j}}+1, \rladder_j]$ is the longest one 
that scores above the cutoff, and, thus Line~\ref{line:seg.test}
finds the optimal segment if the invariant is true.
In order to see that the invariant is true, 
notice the following. 
First, after the condition of the loop in
Line~\ref{line:skip.left} fails with
$j=m$, the invariant  holds by Definition~\ref{def:pairs}
of~$\lpair{m}$. 
Secondly, for $j<m$, 
$\lpair{j}$ 
can be looked for starting the search at $\lpair{j+1}$ 
by Lemma~\ref{lm:pairs.ordered}, and, thus 
the invariant holds every time the execution arrives
to Line~\ref{line:test.left}.
\end{proof}

\section{Related problems}
The same technique applies also to the problem of finding 
a segment with maximum score with a lower bound on the segment length. 
(Albeit Xiaoqiu Huang's algorithm~\cite{subsum.Huang} 
is arguably simpler.)
The idea is to define the left and right pairs by 
thresholding on the segment length 
and then select the one segment with the highest score.

The described algorithm can also be used to find the longest segment
with an average score above a given threshold~$\beta$.
Since $\frac{a_i+a_{i+1}+\dotsc+a_j}{j-i+1}\ge \beta$
if and only if $\sum_{k=i}^j (a_k-\beta)\ge 0$, 
the longest such segment can be found by using 
Algorithm~\Algo{LongestSegment} with scores $(a_i-\beta)$ and 
threshold~$\alpha=0$. 

\section{Acknowledgments}
I would like to thank Mihai Pop and Pavel Havlak  
for discussions of shotgun sequence trimming, as well as 
David Jaffe for sending me the Arachne source code.

\paragraph{Remark.}
Kuan-Yu Chen and Kun-Mao Chao's paper~\cite{segment.long} about the same problem   
came out in print while this paper was under review. (So this will 
always remain a preprint.)
Their algorithm works in an 
on-line setting. They also show the reduction 
for finding the shortest segment which scores above a given cutoff. 

%{\small
%\bibliographystyle{abbrv}
%\bibliography{journals_abbrev,segmentation,pgi}
%}

\end{document}